\documentclass[letter]{aa} 
\usepackage{graphicx}
\usepackage{txfonts}
\begin{document}
   \title{Two very nearby ($d$$\sim$5~pc) ultracool brown dwarfs  
          detected by their large proper motions 
          from WISE, 2MASS, and SDSS data\thanks{based on observations with the Large Binocular Telescope (LBT)}}


\titlerunning{Two very nearby ($d$$\sim$5~pc) ultracool brown dwarfs}

   \author{R.-D. Scholz
          \and
          G. Bihain
          \and
          O. Schnurr
          \and
          J. Storm
          }

   \institute{Leibniz-Institut f\"ur Astrophysik Potsdam (AIP),
              An der Sternwarte 16, 14482 Potsdam, Germany\\
              \email{rdscholz@aip.de, gbihain@aip.de, 
                     oschnurr@aip.de, jstorm@aip.de}
             }

   \date{Received May 20, 2011; accepted July 1, 2011}

 
  \abstract
   {}
   {WISE provides an infrared all-sky survey which aims at completing
    our knowledge on the possibly dramatically increasing number of brown dwarfs
    with lower temperatures. We search for the nearest representatives of the coolest
    brown dwarfs, which will be very interesting for detailed follow-up observations,
    once they haven been discovered.}
   {We have used the preliminary data release from WISE, selected bright
    candidates with colours typical of late-T dwarfs, tried to match them
    with faint 2MASS and SDSS objects, to determine their proper motions,
    and to follow-up them spectroscopically.}
   {We have identified two new ultracool brown dwarfs,
    WISE~J0254$+$0223 and WISE~J1741$+$2553,
    with large proper motions of about 2.5 and 1.5 arcsec/yr, respectively. 
    With their $w1$$-$$w2$$\sim$3.0 and $J$$-$$w2$$\sim$4.0 colour indices
    we expect both to have a spectral type of $\sim$T8-T10 and 
    absolute magnitude of $M_{w2}$$\sim$14. 
    We confirm WISE~J1741$+$2553 as a 
    T9-T10 dwarf from near-infrared spectroscopy with LBT/LUCIFER1.
    From their bright WISE $w2$ magnitudes of 12.7 and 12.3, we estimate
    distances of 5.5$^{+2.3}_{-1.6}$~pc and 4.6$^{+1.2}_{-1.0}$~pc 
    and tangential velocities of $\sim$65~km/s and $\sim$34~km/s 
    indicating Galactic thick and thin disk membership, respectively.}
   {}


   \maketitle
%

\section{Introduction}\label{Sect1}

The Wide-field Infrared Survey Explorer 
(WISE; Wright et al.~\cite{wright10})
observed the sky in four infrared bands
($w1$ at 3.4~$\mu$m, $w2$ at 4.6~$\mu$m, $w3$ at 12~$\mu$m, and $w4$ at 22~$\mu$m).
It allows the detection of nearby cool brown dwarfs (spectral types $>$T5) with much 
higher efficiency than the existing Two Micron All Sky Survey 
(2MASS; Skrutskie et al.~\cite{skrutskie06})  
and the ongoing Sloan Digital Sky Survey with its two recent data releases
(SDSS DR7; Abazajian et al.~\cite{abazajian09},
and SDSS DR8; Aihara et al.~\cite{aihara11}).

\begin{figure}
   \centering
   \includegraphics[width=51mm,angle=-90]{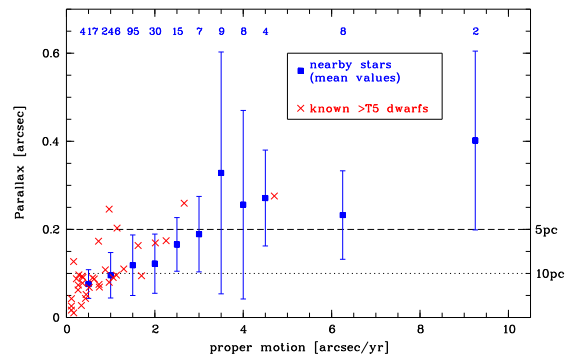}
      \caption{Mean distances of nearby stars (from Gliese \& Jahrei{\ss}~\cite{gliese91})
               with 10\% accurate parallaxes 
               vs. their proper motions. The number of stars
               in each proper motion bin is given on the top. The last two bins 
               are wider to have enough stars per bin. Error bars show
               the width of the distribution in each bin. Known $>$T5 dwarfs
               with parallaxes (from the compilation of Gelino et al.~\cite{gelino11}, 
               supple\-mented by data from Marocco et al.~\cite{marocco10},
               Liu et al.~\cite{liu11}, Luhman, Burgasser \& Bochanski~\cite{luhman11},
               and Subasavage et al.~\cite{subasavage09}) are overplotted as red crosses.
              }
         \label{Fig1}
   \end{figure}

As for the nearest stars in the catalogue of Gliese \& Jahrei{\ss}~(\cite{gliese91})
we expect the nearest cool brown dwarfs to be high proper motion (HPM) objects.
Fig.~\ref{Fig1} shows the correlation between proper motion and parallax. 
HPM stars with $\mu$$>$1~arcsec/yr lie typically within 10~pc. Compared to 
912 nearby stars shown with their mean values,
only few (34) late-T dwarfs with parallax measurements are available.
However, seven out of ten HPM late-T dwarfs with $\mu$$>$1~arcsec/yr
fall in the 10~pc sample, whereas the other three are very close to the 10~pc horizon.

This motivated our HPM search for cool brown dwarfs
in the immediate solar neighbourhood using the preliminary
WISE data release combined with the previous near-infrared (2MASS) 
and deep optical (SDSS) surveys. First results are presented.

\section{Selection of candidates and cross-identification}

To identify possible HPM and thus nearby 
cool brown dwarfs in the WISE data we used the following selection criteria:

\begin{itemize}
\item $w1$$-$$w2$$>$2 and $w2$$-$$w3$$<$2.5 - these are the colour constraints 
also applied by Burgasser et al.~(\cite{burgasser11}) in search of late-T dwarfs, 
where the second condition 
aims at excluding extragalactic sources (Wright et al.~\cite{wright10}).
\item $w2$$<$13 - we considered only bright WISE sources, since those have a higher
probability to show up in 2MASS or SDSS if we consider the limiting magnitudes of these
surveys (see Table 9 in Metchev et al.~\cite{metchev08}) 
and the typical colours of known late-T dwarfs ($J$$-$$w2$$\gtrsim$2;
Mainzer et al.~\cite{mainzer11} and $z$$-$$J$$\sim$3.5; Metchev et al.~\cite{metchev08})
\item We included only WISE point sources outside the Galactic plane ($|b|$$>$10$\degr$
to avoid problems due to reddening.
\item We looked only for WISE sources without a 2MASS counterpart within 3~arcsec,
which implies a minimum proper motion of $\mu$$\gtrsim$0.3~arcsec/yr with the typical
epoch difference of 10 years between 2MASS and WISE.
\end{itemize}

\begin{figure}
   \centering
   \sidecaption
   \includegraphics[width=48mm]{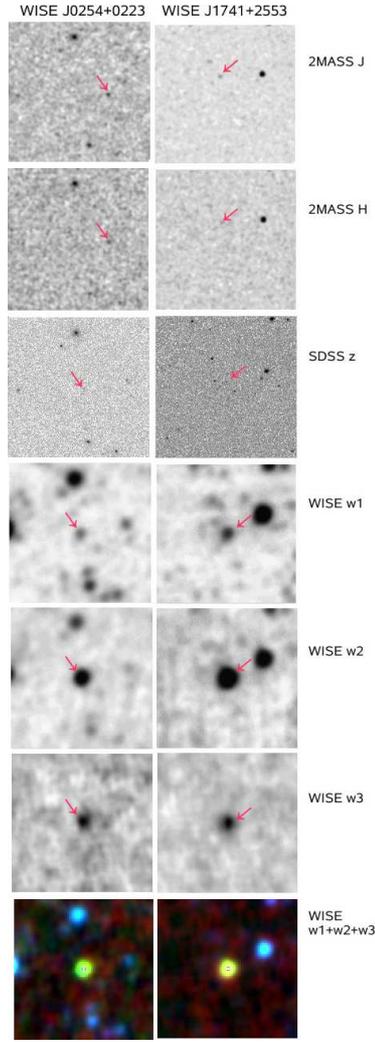}
      \caption{Finding charts of WISE~J0254$+$0223 (left column) and 
               WISE~J1741$+$2553 (right column)
               showing 2$\times$2 arcmin fields centered on the WISE positions
               with North up and East to the left. The charts are
               sorted by the epochs of the images, 2MASS, SDSS, WISE (from top to
               bottom). The lower row shows WISE false colour composites.
              }
         \label{Fig2}
   \end{figure}

Only 98 candidates remained after this selection, and all of them were
subsequently inspected by eye on 2$\times$2 arcmin finding charts 
(as shown in Fig.~\ref{Fig2}) from WISE, 2MASS, and SDSS (if available) 
in search of shifted counterparts with corresponding proper motions 
up to about 6~arcsec/yr. The majority of
candidates were rejected as reddened stars, background galaxies, and artefacts
(in diffraction spikes of bright stars), and for a few candidates no lucid
explanation could be given. However, 
we found
four known T7-T8 dwarfs 
discovered by Burgasser et al.~(\cite{burgasser02}), 
Looper, Kirkpatrick \& Burgasser~(\cite{looper07}) and
Burgasser aet al.~(\cite{burgasser00}), respectively
(see also
Gelino 
et al.~\cite{gelino11} and references therein; with
proper motions from 1.3 to 2 arcsec/yr), 
2MASSI~J0415195$-$093506 (T8 at 5.7~pc),
2MASSI~J0727182$+$171001 (T7 at 9.1~pc), 
2MASS~J07290002$-$3954043 (T8p), and
Gliese~570D (T7.5 at 5.9~pc) 
and two new objects.
There are further 32 known $\gtrsim$T7 dwarfs in 
Gelino et al.~(\cite{gelino11}), of which however, only
UGPS~J072227.51$-$054031.2 (hereafter UGPS~J0722$-$0540; T10 at 4.1~pc)
in the Galactic plane and three 
fainter (13.9$<$$w2$$<$15.2) and likely more distant T7-T7.5 dwarfs are 
measured as resolved sources by WISE.

\section{Discovery of two new HPM brown dwarfs}

The four known late-T dwarfs, which we identified on the WISE and 2MASS
finding charts with positional separations of about 15 to 25 arcsec, showed up as
bright and unusual 'green' sources in the WISE false colour composites and as
relatively faint (15.3$<$$J$$<$15.9) and
'blue' ($-$0.2$<$$J$$-$$H$$<$$+$0.2 and 0.0$<$$J$$-$$K_s$$<$$+$0.3) 2MASS sources.  
Except for the T8p dwarf, they were still detected in the 2MASS $K_s$-band.
For two new 'green' and 'yellow' WISE sources (see Fig.~\ref{Fig2}), 
WISEPC~J025409.45$+$022359.1 (hereafter WISE~J0254$+$0223) and 
WISEPC~J174124.25$+$255319.5 (hereafter WISE~J1741$+$2553), we found
even fainter (16.4$<$$J$$<$16.6) 2MASS counterparts at similar large separations.
The latter two objects are not detected in the 2MASS $K_s$-band, and only 
the second one is blue ($J$$-$$H$$\approx$$+$0.1 and $J$$-$$K_s$$\lesssim$$-$0.3), 
whereas the first one appears to be relatively red 
($J$$-$$H$$\approx$$+$0.7 and $J$$-$$K_s$$\lesssim$$+$0.6). However, the 2MASS
magnitudes of the two new objects are close to the survey limits, and the
corresponding colours from 2MASS are rather uncertain, compared to those of
the four brighter known late-T dwarfs.

\begin{figure}
   \centering
   \includegraphics[width=26mm,angle=-90]{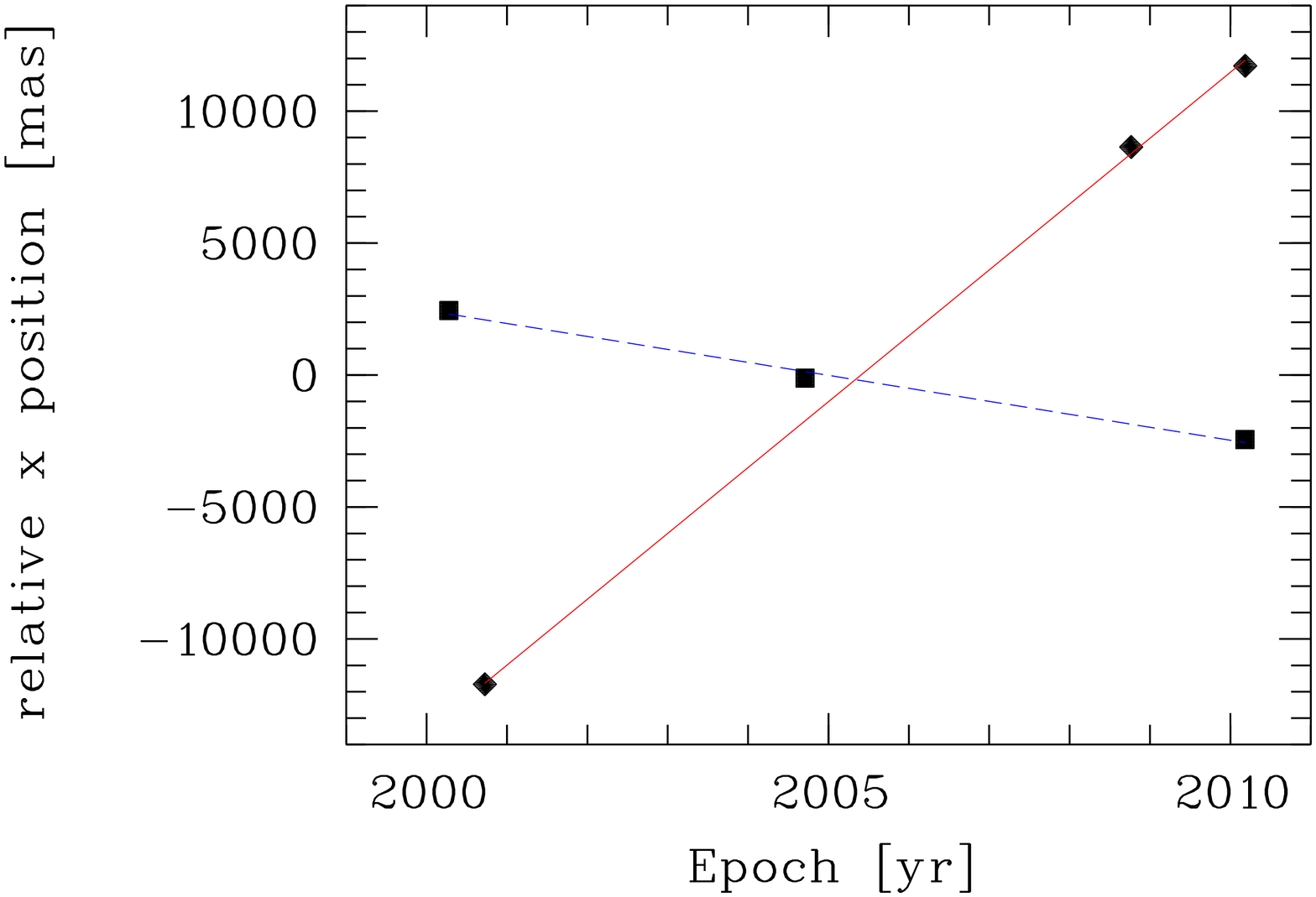}
   \includegraphics[width=26mm,angle=-90]{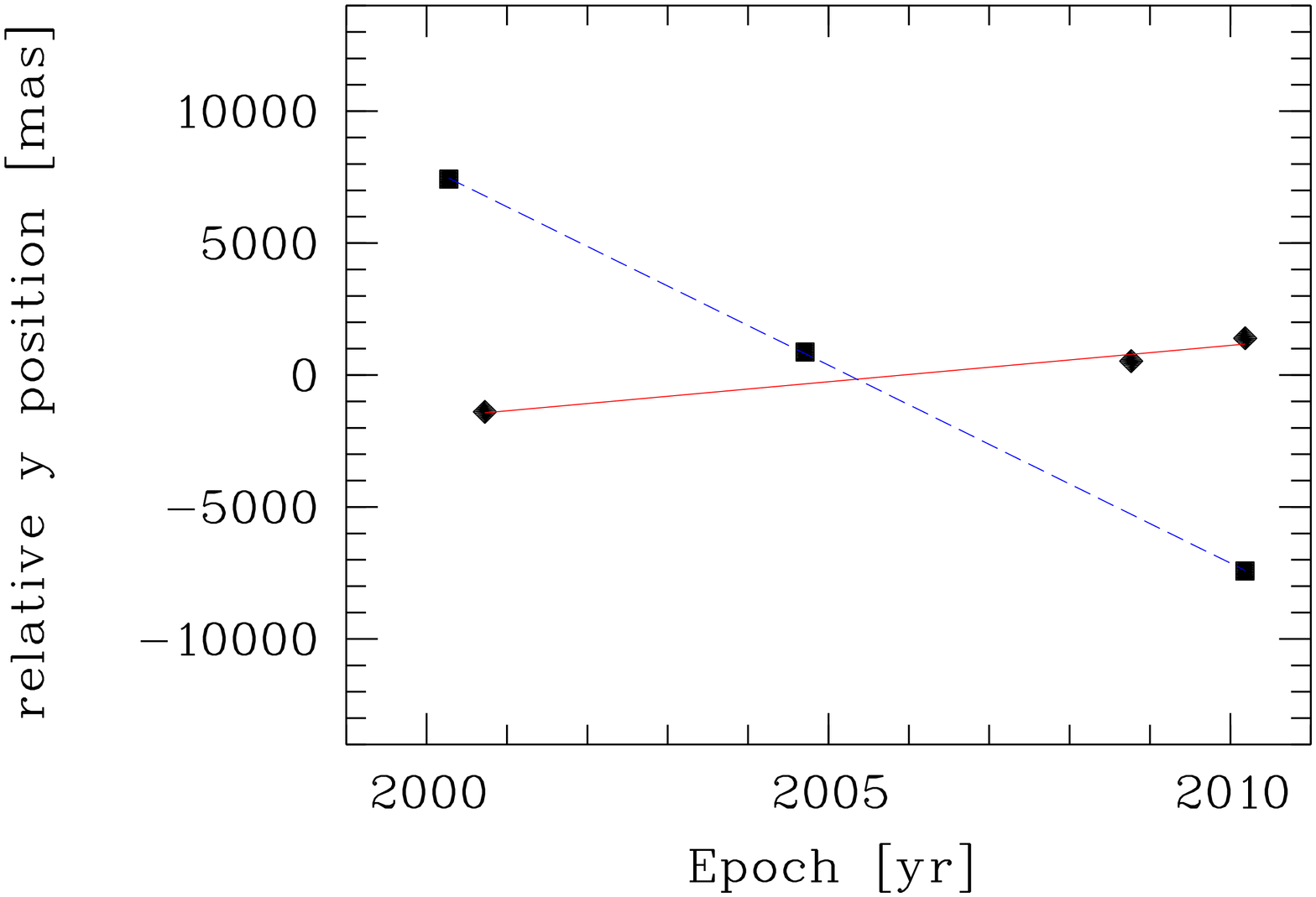}
      \caption{Relative positions from 2MASS, SDSS, and WISE (from left to right)
               and proper motion fits for WISE~J0254$+$0223 (filled lozenges, solid lines)
               and WISE~J1741$+$2553 (filled squares, dashed lines)
              }
         \label{Fig3}
   \end{figure}

The additional identification of WISE~J0254$+$0223 
in SDSS DR8 (Aihara et al.~\cite{aihara11})
and of WISE~J1741$+$2553 in SDSS DR7 (Abazajian et al.~\cite{abazajian09})
led to clear proper motion fits (Fig.~\ref{Fig3}). The very large proper motions
are a first hint that these objects should be very close to the Sun.
Both objects are only detected in the SDSS $z$-band which is typical of nearby
late-T dwarfs. Metchev et al.~(\cite{metchev08}) mentioned that there were
no T8 and later type brown dwarfs known in the SDSS.
The 11 late-T dwarf candidates identified by
Scholz~(\cite{scholz10}) in SDSS and UKIDSS 
(six of which have been confirmed 
spectroscopically as T5-T8 dwarfs by Burningham et al.~\cite{burningham10}, 
Burgasser et al.~\cite{burgasser10}, and Murray et al.~\cite{murray11}) 
have 19.9$<$$z$$<$20.7. None of them were detected in 2MASS.
In comparison, our two new objects are 
with $z$=19.9 and $z$=19.7 relatively bright and have at least 2-3 times larger
proper motions than the above mentioned 11 objects. 

The photometry and
different epoch positions of WISE~J0254$+$0223 and WISE~J1741$+$2553
from SDSS, 2MASS and WISE, as well as their proper 
motions are shown in Table~\ref{table:1}. Note that the proper motions are
derived from simple linear fitting and may be affected by parallactic motions.

%
\begin{table}
\caption{Astrometry \& photometry of two new HPM brown dwarfs}             
\label{table:1}      
\centering                          
\begin{tabular}{l r r}        
\hline\hline                 
Parameter & WISE~J0254$+$0223 & WISE~J1741$+$2553 \\    
\hline                        
WISE RA (J2000)  & 02 54 09.4501   & 17 41 24.2578   \\ 
WISE Dec (J2000) & $+$02 23 59.136 & $+$25 53 19.507 \\
WISE epoch       & 2010.182        & 2010.182        \\
SDSS RA (J2000)  & 02 54 09.2449   & 17 41 24.4306   \\
SDSS Dec (J2000) & $+$02 23 58.268 & $+$25 53 27.805 \\
SDSS epoch       & 2008.764        & 2004.7095       \\
SDSS run/data release & 7717/DR8   & 4832/DR7    \\
2MASS RA (J2000) & 02 54 07.8864   & 17 41 24.6206   \\
2MASS Dec (J2000)& $+$02 23 56.346 & $+$25 53 34.361 \\
2MASS epoch      & 2000.723        & 2000.277        \\
$\mu_{\alpha}\cos{\delta}$ [mas/yr] & $+$2496$\pm$46 &  $-$492$\pm$43 \\
$\mu_{\delta}$ [mas/yr]             &  $+$276$\pm$47 & $-$1500$\pm$11 \\
SDSS $z$ [mag]   & 19.861$\pm$0.074 & 19.745$\pm$0.105 \\
2MASS $J$ [mag]  & 16.557$\pm$0.156 & 16.451$\pm$0.099 \\
2MASS $H$ [mag]  & 15.884$\pm$0.199 & 16.356$\pm$0.216 \\
2MASS $K_s$ [mag]& $>$16.006~~~~~   & $>$16.785~~~~~ \\
WISE $w1$ [mag]  & 15.743$\pm$0.070 & 15.228$\pm$0.040 \\
WISE $w2$ [mag]  & 12.707$\pm$0.031 & 12.312$\pm$0.025 \\
WISE $w3$ [mag]  & 11.042$\pm$0.131 & 10.675$\pm$0.075 \\
\hline                                   
\end{tabular}
\tablefoot{We have also detected both objects at the positions
02 54 07.301 $+$02 23 56.91 (1996.805) and
17 41 24.649 $+$25 53 40.25 (1996.449), respectively
on $I$-band photographic Schmidt plates (not shown in Fig.~\ref{Fig2})
scanned in the
SuperCOSMOS Sky Surveys (Hambly et al.~\cite{hambly01}).
The proper motions involving these positions 
($+$2438.7$\pm$40.6, $+$176.1$\pm$64.1 and
$-$407.8$\pm$59.7, $-$1507.2$\pm$7.0, respectively) confirm
our HPM measurements but
are slightly less accurate than our preferred solutions
given in the table above and shown in Fig.~\ref{Fig3}.}
\end{table}

\section{Photometric classification and distances}

With their large $w1$$-$$w2$ colour indices of $+$3.04 and $+$2.92 
WISE~J0254$+$0223 and WISE~J1741$+$2553 are expected
to have spectral types of $\sim$T7-T10 according to 
Mainzer at al.~(\cite{mainzer11}),
although we find for the four known T7-T8 dwarfs (two of which are
listed in Mainzer at al.~\cite{mainzer11} with different WISE photometry)
detected in our HPM search 
(Sect.~\ref{Sect1}) somewhat smaller values of $+$2.30$<$$w1$$-$$w2$$<$$+$2.85. 
Therefore, we believe our two new objects are more likely 
ultracool ($\gtrsim$T8) brown dwarfs.
In support of this classification we mention their large
$J$$-$$w2$ indices of $+$3.85 and $+$4.14 (compared to values from $+$2.67 to $+$3.46 
for the four known T7-T8 dwarfs) corresponding to
an absolute magnitude of $M_{w2}$$\sim$14 if we use the 
linear relation between $M_{w2}$ and $J$$-$$w2$ given in Wright et al.~(\cite{wright11}).
For comparison, UGPS~J0722$-$0540 
(Lucas et al.~\cite{lucas10}), is detected in WISE with $w2$=12.17
and has an absolute magnitude of $M_{w2}$=14.11, whereas all T dwarfs with known
distances from Patten et al.~(\cite{patten06}) plotted by Wright et al.~(\cite{wright11})
have brighter absolute magnitudes ($M_{w2}$$<$13.5) with colour indices $J$$-$$w2$$<$$+$3.3. 
The $J$$-$$w2$ colours of
WISE~J0254$+$0223 and WISE~J1741$+$2553 do not reach the very large values ($\gtrsim$5)
of the two coolest known (probably $>$T10) 
but more distant brown dwarfs WD~0806$-$661B at $\sim$19~pc 
(Luhman, Burgasser \& Bochanski~\cite{luhman11}; Rodriguez et al.~\cite{rodriguez11};
Subasavage et al.~\cite{subasavage09}) 
and CFBDSIR J145829$+$101343B at $\sim$23~pc (Liu et al.~\cite{liu11}) 
but are comparable with that of the A component (T9.5) of the latter as shown in
Wright et al.~(\cite{wright11}).

For a first approximation of the distances of WISE~J0254$+$0223 and WISE~J1741$+$2553,
we use the preliminary classification of both objects as $\sim$T8-T10 dwarfs with
an absolute magnitude of $M_{w2}$$\sim$14. 
Conservatively, we assume an uncertainty
of 0.75~mag in this absolute magnitude, taking into account
the $\sim$0.5~mag spread in absolute magnitudes shown in
Wright et al.~(\cite{wright11}) and the $J$$-$$w2$ colour spread 
observed for the known T8-T10 dwarfs.
There is also a significant scatter in near-infrared
(MKO) to mid-infrared (Spitzer) colours of late-T dwarfs
(Leggett et al.~\cite{leggett10}).
We then arrive at 
photometric distances from the Sun for WISE~J0254$+$0223 and 
WISE~J1741$+$2553 of 5.5$^{+2.3}_{-1.6}$~pc and 4.6$^{+1.9}_{-1.4}$~pc, 
respectively.

With these distance estimates, we compute absolute magnitudes $M_z$$\sim$21.2 and 
$M_z$$\sim$21.4, respectively for WISE~J0254$+$0223 and WISE~J1741$+$2553. 
These values are 1.3-1.5~mag
fainter than the $M_z$$\sim$19.9 of 
Ross~458C 
(=Hip 63510C), the only 
$\gtrsim$T8 
dwarf with 
distance and SDSS detection now available 
(Goldman et al.~\cite{goldman10}; 
Scholz~\cite{scholz10}; Burgasser et al.~\cite{burgasser10}) The latter is 
in good agreement with
the assumed $M_z$$\sim$20.0 for T8 dwarfs in Metchev et al.~(\cite{metchev08}).
On the other hand, the $z$-band photometry obtained by Lucas et al.~(\cite{lucas10})
for the T10 dwarf UGPS~J0722$-$0540 leads to $M_z$ values between 22.4 and 22.6,
about 1~mag fainter than our values. 

   \begin{figure}
   \centering
   \includegraphics[width=53mm, angle=-90]{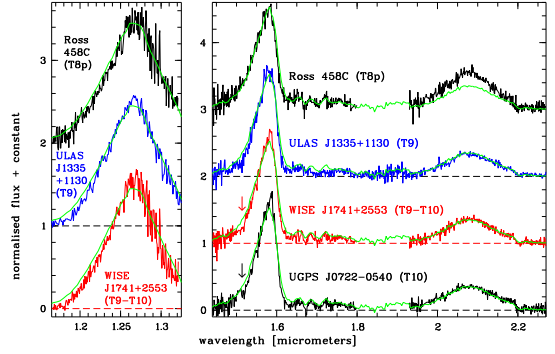}
      \caption{LBT/LUCIFER1 $J$- (left) and 
               $H$$+$$K$-band  
               (right) 
               spectra of WISE~J1741$+$2553 and 
               two other objects
               (see Table~\ref{table:2}).
               Also shown is a T9 template spectrum
               from Burningham et al.~(\cite{burningham08}). 
               A lower-resolution spectrum of the 
               T8 template 2MASSI~J0415195$-$093506 (from
               Burgasser et al.~\cite{burgasser04}) is overplotted.
               Arrows mark the candidate NH$_3$ absorption feature,
               mentioned by Lucas et al.~(\cite{lucas10}).
              }
         \label{Fig4}
   \end{figure}

%
\begin{table*}
\caption{Spectral indices obtained from LBT/LUCIFER1 spectra (SpT from comparison with T9 average from Lucas et al.~\cite{lucas10} in brackets)}
\label{table:2}
\centering
\begin{tabular}{l c c c c c c c }   
\hline\hline
Object            & SpT & Reference for SpT                             & W$_J$         & NH$_3$ - H   & H$_2$O -H    & CH$_4$ - H   & CH$_4$ - K \\
\hline
Ross~458C         & T8p  & Burgasser et al.~\cite{burgasser10}  & 0.264 ($<$T9) & 0.664 ($<$T9)& 0.206 ($<$T9)& 0.107 ($<$T9)& 0.054 ($>$T9) \\
UGPS~J0722$-$0540 & T10 & Lucas et al.~\cite{lucas10}                   &               & 0.530 ($>$T9)& 0.141 ($<$T9)& 0.055 ($>$T9)& 0.079 ($>$T9) \\
WISE~J1741$+$2553 & T9-T10 & this paper                        & 0.234 ($>$T9) & 0.551 ($<$T9)& 0.119 ($>$T9)& 0.066 ($>$T9)& 0.056 ($>$T9) \\
\hline
\end{tabular}
\tablefoot{In the T dwarf 
classification system originally defined by
Burgasser et al.~(\cite{burgasser06}) the CH$_4$ - K index was shown to be
degenerate for types $>$T6. Extending the system
to T9 Burningham et al.~(\cite{burningham08}) showed that the CH$_4$ - H 
index is also degenerate with $>$T7, whereas the T8, T9 average, and T10
values of the latter show a clear trend in Lucas et al.~(\cite{lucas10}).
We consider Ross~458C as T8p since it 
is a relatively young (0.15-0.80 Gyr) 
object (Burgasser et al.~\cite{burgasser10}), whereas the field brown dwarfs are 
assumed to be older (0.2-10~Gyr; see e.g. Lucas et al.~\cite{lucas10}).}
\end{table*}

\section{Spectroscopic observations with LBT/LUCIFER}

We have used the Large Binocular Telescope (LBT) near-infrared spectrograph LUCIFER1
(Mandel et al.~\cite{mandel08}; Seifert et al.~\cite{seifert10}; 
Ageorges et al.~\cite{ageorges10}) in long slit 
spectroscopic mode with 
the $H$$+$$K$ (200 lines/mm + order separation filter) and $zJHK$ gratings 
(210 lines/mm + $J$ filter) 
to observe UGPS~J0722$-$0540 on 2011 March 06 (only in $H$$+$$K$, total 
integration 30~min) and WISE~J1741$+$2553 (40~min $H$$+$$K$, 27~min $J$) 
together with 
Ross~458C 
(30~min $H$$+$$K$, 20~min $J$) on 2011 May 12. 
The central wavelengths were chosen at 1.835~$\mu$m ($H$$+$$K$) and
1.25~$\mu$m ($J$) yielding a coverage of 1.38--2.26 and
1.17--1.32~$\mu$m, respectively.
The slit width was of 2~arcsec for UGPS~J0722$-$0540 and 1~arcsec (4 pixels) 
for the other objects. For the 1~arcsec slit, the spectral resolving power is
$R$=$\lambda$/$\Delta$$\lambda$$\approx$4230, 940, and 1290 at $\lambda$$\approx$1.24,
1.65, and 2.2~$\mu$m, respectively. Observations consisted of individual
75~s exposures in $H$$+$$K$ and 200~s exposures in $J$ with shifting the target along
the slit following an ABBA pattern (to allow sky subtraction)
until the total integration time was reached.
A0V standards were observed just before or after the targets with 
similar airmass.

The raw spectroscopic data were reduced using standard routines within the {\tt
IRAF\footnote{IRAF is distributed by the National Optical Astronomy
Observatories, which are operated by the Association of Universities for
Research in Astronomy, Inc., under cooperative agreement with the National
Science Foundation.}} environment. The spectra were sky-subtracted, aligned,
combined, optimally extracted, and wavelength-calibrated using vacuum
wavelengths of Ar arc lamps and the deep telluric absorption features. No flat
division was applied to avoid including additional noise features from the
halogen lamp. The instrumental response and telluric bands were removed by
dividing by the A0 star spectra and multiplying by a black-body spectrum with
the same effective temperature. The intrinsic lines in the hot-star spectra were
removed before dividing the science spectra. Residuals from intense sky
emission lines in the target spectra were removed by interpolation
across the lines. The $J$ and $H$$+$$K$ spectra were normalised by the 
average flux in the range 1.20--1.30 and 1.52--1.61~$\mu$m, respectively.

The $H$$+$$K$ 
spectra of WISE~J1741$+$2553 and the T10 dwarf
UGPS~J0722$-$0540 are very similar, including a possible common
NH$_3$ feature expected for Y-type objects, 
whereas both spectra ($H$$+$$K$ and $J$) 
of WISE~J1741$+$2553 
indicate a type later than T8
 (Fig.~\ref{Fig4}). 
We have computed spectral indices used for the classification of ultracool brown dwarfs
(see Burningham et al.~\cite{burningham08} and references therein) and compare them with
average values of T9 dwarfs given by Lucas et al.~(\cite{lucas10}) in Table~\ref{table:2}.
Unfortunately, no other indices than W$_J$ could be determined in the $J$-band, since
the LUCIFER $J$ grating provides a very narrow wavelength interval. Our indices 
are in good agreement with the measurements of Lucas et al.~(\cite{lucas10})
for UGPS~J0722$-$0540 and of Burgasser et al.~\cite{burgasser10} for 
Ross~458C,
except for NH$_3$ - H, where our indices are larger by 
about 0.04 and 0.06,
respectively.  
As in the case of the known T10 dwarf, all but one indices of WISE~J1741$+$2553 
classify it as $>$T9,
although the usefulness of some indices has been questioned (see Notes
to Table~\ref{table:2}). 
We assign conservatively
a spectral type of 
T9-T10, reduce the assumed absolute magnitude uncertainty to 0.5~mag, 
and get
a more accurate spectroscopic/photometric distance of 4.6$^{+1.2}_{-1.0}$~pc.

\section{Conclusions}

   \begin{enumerate}
      \item While WISE~J0254$+$0223 and WISE~J1741$+$2553 are likely 
            similar to the few
            other T8-T10 brown dwarfs known, they are the first 
            ultracool brown dwarfs detected in both 2MASS and SDSS. 
            With their relatively
            bright magnitudes they are excellent targets for detailed
            spectroscopic investigations and for high resolution imaging in search
            of possible binarity. They may become important 
            laboratory 
            sources at
            the boundary between the T-type and the suggested Y-type 
            (Kirkpatrick et al.~\cite{kirkpatrick99})
            classes of brown dwarfs.
      \item Our initial 
            photometric classification of both objects as 
            $\sim$T8-T10 dwarfs
            was confirmed for WISE~J1741$+$2553, for which we obtained 
            spectroscopy with LBT/LUCIFER1 and classified it  
            more accurately as a T9-T10 dwarf.
      \item Our 
            photometric(/spectroscopic) distances 
            place both objects at about 5~pc from
            the Sun, probably nearly as close as the T10 dwarf UGPS~J0722$-$0540
            in the Galactic plane ($b$=$+$4$\degr$),
            and make them highly interesting targets for near-infrared parallax
            programs. 
            With the upper limits of our distance estimates
            they fall still clearly in the 10~pc sample.
      \item The larger proper motion and slightly larger distance of
            the higher Galactic latitude object WISE~J0254$+$0223 ($b$=$-$48$\degr$)
            lead to a relatively high tangential velocity 
            of $\sim$65~km/s (possibly indicating a Galactic thick disk membership), 
            compared to $\sim$34~km/s for WISE~J1741$+$2553 ($b$=$+$26$\degr$), 
            which is typical of the thin disk population.
   \end{enumerate}

\begin{acknowledgements}
The authors would like to thank Roland Gredel, Jochen Heidt, Jaron Kurk, Ric Davies,
and all observers at the LBT for their assistance and help during the
preparation and execution of the LUCIFER observations,
Adam Burgasser and Ben Burningham for providing T8/T9 template spectra 
and the anonymous referee for her/his helpful comments and suggestions.

This research has made use of the
NASA/IPAC Infrared Science Archive, which is operated by the Jet Propulsion
Laboratory, California Institute of Technology, under contract with the
National Aeronautics and Space Administration,
and of data products from WISE,
which is a joint project of the University of California, 
Los Angeles, and the Jet Propulsion Laboratory/California Institute of 
Technology, funded by the National Aeronautics and Space Administration,
from 2MASS, and from SDSS DR7 and DR8. 
Funding for SDSS-III has been provided by the Alfred P. Sloan Foundation, 
the Participating Institutions, the National Science Foundation, and the 
U.S. Department of Energy. The SDSS-III web site is http://www.sdss3.org/.

\end{acknowledgements}

\end{document}